\documentclass[%
 aip,
 amsmath,amssymb,
 reprint,%
]{revtex4-1}

\usepackage{graphicx}
\usepackage{dcolumn}
\usepackage{bm}
\usepackage[utf8]{inputenc}
\usepackage[T1]{fontenc}
\usepackage{mathptmx}
\usepackage{etoolbox}
\usepackage{lipsum}
\usepackage{float}
\usepackage{siunitx}
\usepackage[english]{babel}
\makeatletter
\def\@email#1#2{%
 \endgroup
 \patchcmd{\titleblock@produce}
  {\frontmatter@RRAPformat}
  {\frontmatter@RRAPformat{\produce@RRAP{*#1\href{mailto:#2}{#2}}}\frontmatter@RRAPformat}
  {}{}
}%
\makeatother

\draft 

\begin{document}


\title[SHIELD: A Reference Gas-Driven Permeation Platform for Hydrogen Permeation Studies]{SHIELD: A Reference Gas-Driven Permeation Platform for Hydrogen Permeation Studies} 


\author{J. Dark}
\email{darkj385@mit.edu}
\author{C. Weaver}
\author{R. Delaporte-Mathurin}
\author{S. Ferry}
\author{K. B. Woller}

\affiliation{Plasma Science and Fusion Center, Massachusetts Institute of Technology, Cambridge, MA 02139, USA}

\date{\today}

\begin{abstract}
A gas-driven permeation (GDP) platform, SHIELD (Salt-compatible Hydrogen barrier Investigation and EvaLuation for fusion Devices), has been developed to measure hydrogen transport properties in structural materials under controlled thermal and pressure conditions. 
The system is designed to minimise experimental uncertainties associated with leaks, temperature instability, and pressure measurement, while providing reproducible conditions for permeation experiments. 
The rig operates in a static GDP configuration with independent upstream and downstream volumes, enabling precise control of driving pressure and accurate measurement of downstream pressure rise. 
An openly documented data acquisition and processing framework is implemented to ensure data traceability and reproducibility. 
The platform's performance is demonstrated by hydrogen permeation measurements on 316 stainless steel and AISI 1018 low-carbon steel over the temperature range of \SIrange{100}{600}{\degreeCelsius}.
Steady-state permeation fluxes are extracted from linear downstream pressure rise and used to determine permeability.
The measured permeability exhibits Arrhenius behaviour and agrees well with published literature data for both materials. 
Permeability measurements are shown to be robust and reproducible.
These results demonstrate that SHIELD provides a reliable reference platform for hydrogen permeation measurements and is well-suited to evaluating permeation barrier coatings and advanced materials for fusion applications.
\end{abstract}


\maketitle

\section{Introduction}
\label{sec:intro}

Hydrogen permeation through structural materials is a critical phenomenon in nuclear fusion, particularly in applications involving tritium management \cite{gilbert_fusion_2024, coleman_demo_2019, sawan_physics_2006}.
Quantitative determination of hydrogen transport properties is essential for material qualification and permeation barrier assessment \cite{forcey_hydrogen_1988, shimada_tritium_2018, perujo_tritium_1995, nemanic_hydrogen_2019}.
Gas-driven permeation (GDP) experiments are widely employed for this purpose, enabling the extraction of transport properties such as diffusivity, solubility, and permeability by imposing a pressure differential across a specimen and monitoring the pressure rise in a control volume due to the transmitted hydrogen flux \cite{montupet-leblond_permeation_2021, xu_transport_2020, uehara_hydrogen_2015}.

Although conceptually straightforward, permeation measurements are experimentally demanding.
Results are highly sensitive to parasitic leaks, surface effects, and temperature stability.
Reliable operation, therefore, requires careful control of sealing interfaces, thermal conditions, pressure regulation, and detection sensitivity \cite{guo_deuterium_2025}.

This paper presents the Salt-compatible Hydrogen barrier Investigation and EvaLuation for fusion Devices (SHIELD) gas-driven permeation rig, a purpose-built experimental platform designed to deliver reproducible hydrogen permeation measurements under controlled thermal and pressure conditions. 
In addition, the platform incorporates an openly documented data acquisition and processing framework to promote transparency and reproducibility \cite{dark_pttepxmitshield_das_2025, dark_shield-data_2025}.
The theoretical background and parameter-extraction methodology are first summarised, followed by a detailed description of the experimental apparatus.
Measurements are performed using pressure-rise experiments on 316 stainless steel and AISI low-carbon steel specimens, from which permeation properties are extracted and benchmarked against established literature data. 
The resulting agreement demonstrates the platform's measurement capability and establishes SHIELD as a reference facility for future permeation and coating evaluation studies.

\section{Hydrogen permeation theory}
\label{sec:permeation_theory}

Hydrogen permeation in metals results from coupled surface exchange and bulk diffusion processes.
Molecular hydrogen dissociates at the entry surface into atomic species, which migrate through the lattice and recombine at the exit surface. 
The overall permeation rate is therefore governed by both surface reaction kinetics and bulk transport properties.

Bulk transport is described by Fick's first law,
\begin{equation}
    J = -D \nabla c,
\end{equation}
where $J$ is the particle flux (\si{m^{-2}.s^{-1}}), $D$ is the diffusion coefficient (\si{m^{2}.s^{-1}}), and $c$ is the hydrogen concentration (\si{m^{-3}}). 
Combining Fick's law with mass conservation yields the diffusion equation,
\begin{equation}
    \frac{\partial c}{\partial t} = \nabla \cdot (D \nabla c),
\end{equation}
which governs the temporal and spatial evolution of mobile hydrogen within the material.

The diffusion coefficient is temperature-dependent and commonly expressed using an Arrhenius relation,
\begin{equation}
    D = D_0 \exp\left(\frac{-E_D}{k_\mathrm{B}T}\right),
\end{equation}
where $D_0$ is the pre-exponential factor, $E_{D}$ is the activation energy for diffusion (\si{eV}), $k_\mathrm{B}$ is the Boltzmann constant (\SI{8.617e-5}{eV.K^{-1}}) and $T$ is the temperature (\si{K}). 
Effects such as stress-assisted diffusion~\cite{reddy_phase-field_2025}, concentration-dependent interactions~\cite{holtslander_recovery_1979}, or thermophoresis (Soret effect)~\cite{longhurst_soret_1985} are not considered in the present work, as the SHIELD platform is designed to operate under conditions where these effects are negligible.

\begin{figure}[H]
    \centering
    \includegraphics[width=0.8\linewidth, trim={11.8cm 2.5cm 11.8cm 3cm}, clip]{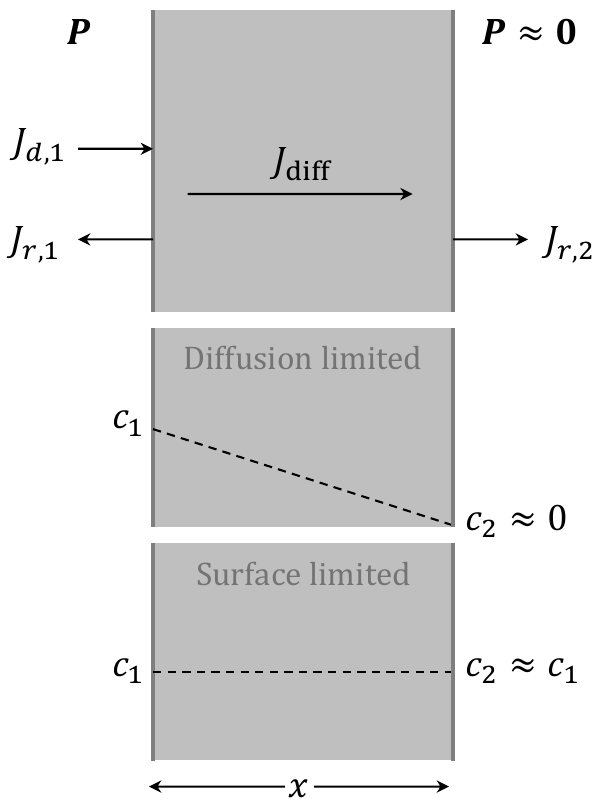}
    \caption{Schematic illustration of hydrogen permeation regimes under $P_{\mathrm{down}} \approx 0$.}
    \label{fig:perm_regimes}
\end{figure}

At the surface, hydrogen solubility arises from the balance between dissociation of molecular hydrogen and recombination of dissolved atomic species.
In typical gas-driven permeation experiments, the downstream pressure is maintained close to zero, $P_{\mathrm{down}} \approx 0$, such that transport is driven by the upstream pressure (see Fig \ref{fig:perm_regimes}).

The flux entering the solid through dissociative adsorption is written as
\begin{equation}
    J_{d,1} = K_{d} P,
\end{equation}
where $K_{d}$ is the dissociation rate constant (\si{m^{-2}.s^{-1}.Pa^{-1}}) and $P$ is the external hydrogen pressure (\si{Pa}). 
The opposing recombination flux is proportional to the square of the mobile concentration,
\begin{equation}
    J_{r,1} = K_r c_1^{2},
\end{equation}
where $K_r$ is the recombination rate constant (\si{m^{4}.s^{-1}}).
Both surface rate constants follow Arrhenius laws,
\begin{equation}
    K_d = K_{d,0} \exp\left(\frac{-E_{kd}}{k_\mathrm{B}T}\right), \quad
    K_r = K_{r,0} \exp\left(\frac{-E_{kr}}{k_\mathrm{B}T}\right).
\end{equation}
Depending on the relative rates of surface reactions and bulk diffusion, two limiting regimes can be identified.

At the upstream surface, the following flux balance applies:
\begin{equation}
    J_{d,1} - J_{r,1} = J_{\mathrm{diff}}.
\end{equation}
However, when $J_{\mathrm{diff}} \ll J_{d,1} - J_{r,1}$:
\begin{equation}
    J_{d,1} \approx J_{r,1},
\end{equation}
which leads to:
\begin{equation}
    K_{d} P = K_r c^{2},
\end{equation}
which in turn gives rise to Sievert's law,
\begin{equation}
    c = K_s\sqrt{P}, \quad \mathrm{with} \quad  K_s = \sqrt{\frac{K_d}{K_r}}.
\end{equation}
Where $K_s$ is the material solubility (\si{m^{-3}.Pa^{-0.5}}).
Under these conditions, transport is \textit{diffusion-limited} and the steady state flux is governed by the concentration gradient across the specimen.
For a sample of thickness $x$, this yields
\begin{equation}
    J_{DL} = \frac{D\,K_s \sqrt{P}}{x}.
\end{equation}

In the opposite limit, where bulk diffusion is much faster than surface reactions, $J_{\mathrm{diff}} \gg J_{d,1} - J_{r,1}$, the diffusive flux is negligible,
\begin{equation}
    J_{\mathrm{diff}} = -D \frac{c_2 - c_1}{x} \approx 0.
\end{equation}
Since $D \neq 0$ and $x \neq 0$, this implies a uniform concentration throughout the specimen (see Fig~\ref{fig:perm_regimes}),
\begin{equation}
    c_1 \approx c_2.
\end{equation}
The value of the uniform concentration can then be determined from the flux balance across the sample,
\begin{equation}
    J_{d,1} - J_{r,1} = J_{r,2}.
    \label{eq: flux balance surfaces}
\end{equation}
Since $c_1 \approx c_2$, the recombination fluxes are equal, $J_{r,1} \approx J_{r,2}$.
Therefore
\begin{equation}
    J_{d,1} = 2J_{r}.
    \label{eq: flux balance surfaces 2}
\end{equation}
This yields a uniform concentration given by: 
\begin{equation}
    c_2 \approx c_1 = \frac{1}{\sqrt{2}} \sqrt{\frac{K_d}{K_r} \ P} = \frac{1}{\sqrt{2}} K_s \sqrt{P},
\end{equation}
which corresponds to a modified Sievert's relation with a factor $1/\sqrt{2}$ arising from the surface flux balance.
The permeation flux, equal to the downstream recombination flux, is obtained from the same flux balance Eq. \eqref{eq: flux balance surfaces 2}:
\begin{equation}
    J_{d,1} = 2J_r = 2J_{SL},
\end{equation}
leading to:
\begin{equation}
    J_{SL} = \frac{1}{2}K_d P,
\end{equation}
indicating that permeation is limited by dissociative adsorption.

Experimentally, the governing transport regime can be identified from the pressure dependence of the permeation flux.
In the \textit{diffusion-limited} regime, $J_{DL}$, the flux scales with the square root of the upstream pressure,
\begin{equation}
    J_{DL} \propto \sqrt{P}.
\end{equation}
In contrast, in the \textit{surface-limited} regime, $J_{SL}$, the permeation flux is proportional to the pressure,
\begin{equation}
    J_{SL} \propto P.
\end{equation}
Therefore, analysis of the scaling of the measured flux with pressure provides a simple and robust experimental criterion to distinguish between \textit{diffusion-limited} and \textit{surface-limited}.

More generally, hydrogen permeation is described by a flux balance at both surfaces coupled with diffusion through the bulk.
Normalising this general solution by the \textit{diffusion-limited} flux leads to a dimensionless formulation, from which a characteristic parameter naturally emerges,
\begin{equation}
    W = \frac{K_d\,x\, \sqrt{P_\mathrm{up}}}{D\,K_s}.
\end{equation}
This permeation number, $W$, quantifies the relative importance of surface kinetics and bulk diffusion.
For $W \gg 1$, the system operates in the \textit{diffusion-limited} regime, whereas for $W \ll 1$, surface kinetics dominate in the \textit{surface-limited} regime.

\subsection{Permeability evaluation from pressure rise}

The SHIELD platform operates as a gas-driven permeation system with a sealed downstream volume of a fixed size.
In contrast to other permeation platforms that rely on continuous pumping and mass spectrometry for flux measurement, the downstream section is isolated and maintained leak-tight.
Hydrogen permeating through the specimen accumulates in this volume, and the resulting increase in pressure is directly monitored.

\begin{figure}[H]
    \centering
    \includegraphics[width=\linewidth, trim={0.5cm 0.5cm 0.25cm 0.5cm}, clip]{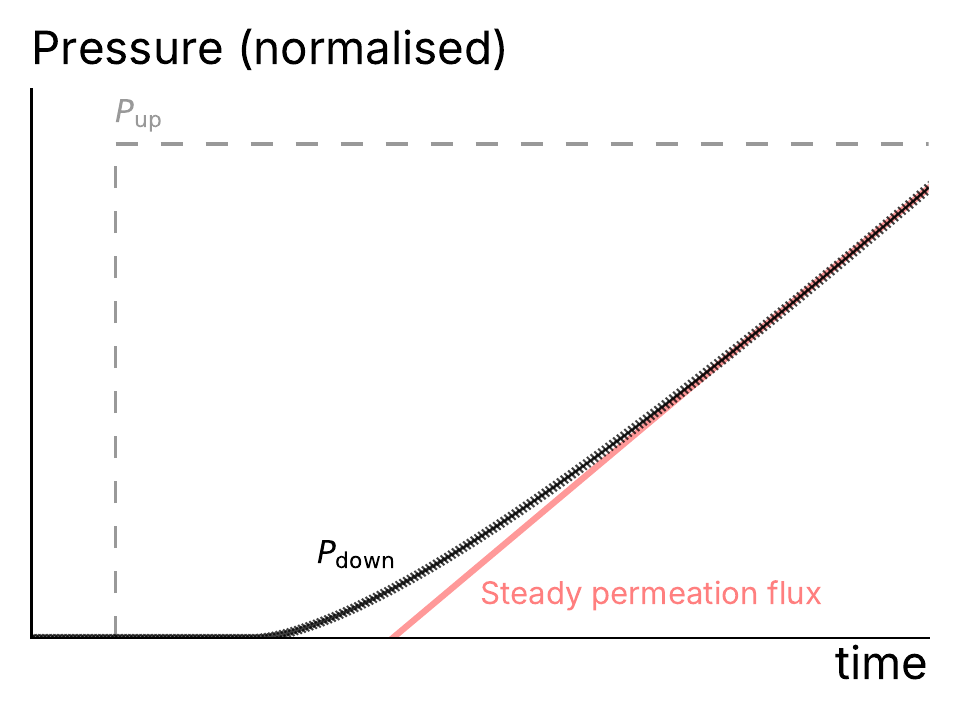}
    \caption{Schematic evolution of downstream pressure following application of an upstream pressure step. After an initial transient, a linear increase in downstream pressure is observed, corresponding to steady-state permeation. The slope of the linear regime is directly proportional to the permeation flux.}
    \label{fig:pressure_rise_ex}
\end{figure}

This approach enables the permeation flux to be inferred from the rate of the pressure rise, provided that the system is leak-tight and that contributions from residual gases are negligible.
Upon application of an upstream pressure step, hydrogen permeates through the specimen and accumulates in the downstream volume.
After an initial transient, a steady-state regime is reached in which the downstream pressure increases linearly with time, corresponding to a constant permeation flux (see Fig~\ref{fig:pressure_rise_ex}).

At steady-state, the rate of particle accumulation in the downstream volume is
\begin{equation}
    \frac{\delta N}{\delta t} = J\,A,
\end{equation}
where $A$ is the exposed sample area (\si{m^{2}}).
Using the ideal gas law, $PV = nRT$, the corresponding pressure rise rate becomes:
\begin{equation}
     \frac{\delta P}{\delta t} = J\,R\,T\, \frac{A}{V},
     \label{eq:pressure-rise}
\end{equation}
where $V$ is the downstream volume (\si{m^3}) and $R=\SI{8.314}{J.mol^{-1}.K^{-1}}$ is the gas constant.

The permeation flux is therefore directly obtained from the measured pressure rise rate.
Under \textit{diffusion-limited} conditions with $P_{\mathrm{down}} \approx 0$, the permeability is given by
\begin{equation}
    \Phi = \frac{J\,x}{\sqrt{P_{\mathrm{up}}}},
\end{equation}
ensuring that the measured flux reflects the intrinsic material permeability of the specimen.

\section{Experimental setup: the SHIELD GDP rig}
\label{sec:setup}

\begin{figure*}[ht!]
    \centering
    \includegraphics[width=0.75\linewidth]{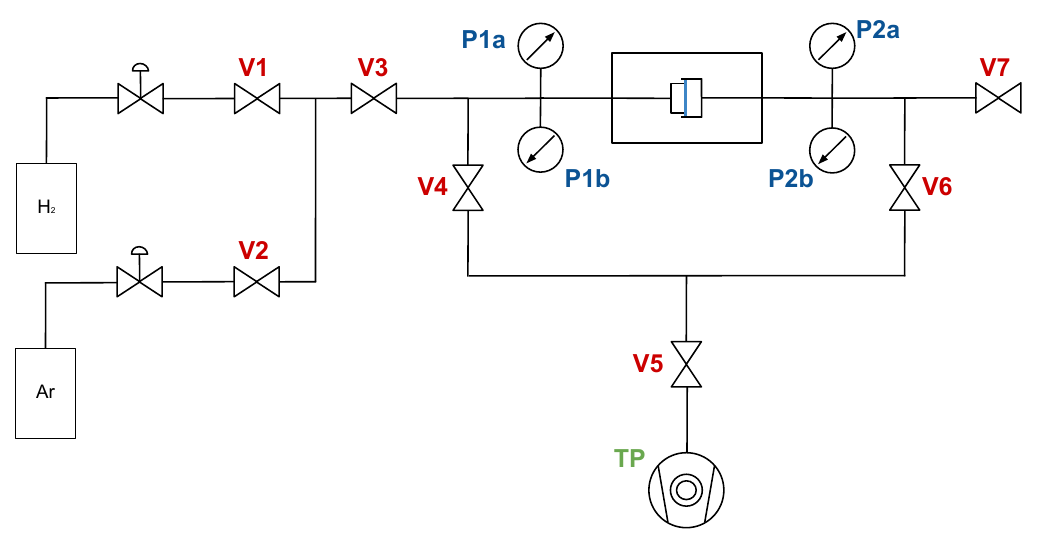}
    \caption{Schematic representation of the PI\&D diagram of the SHIELD experimental platform}
    \label{fig:rig_schematic}
\end{figure*}

The SHIELD Gas Driven Permeation (GDP) platform was developed to measure hydrogen permeation through metallic samples under controlled pressure and temperature conditions.
The system is configured in a static GDP arrangement.
After evacuation and thermal stabilisation, hydrogen is introduced to the upstream volume to a defined pressure, and the section is isolated.
The downstream volume is sealed, and the resulting pressure rise due to permeation through the sample is recorded over time.

SHIELD is designed to operate over a temperature range of \SI{100}{\degreeCelsius} to \SI{600}{\degreeCelsius}.
Base pressures on the order of \SI{e-6}{Torr} are routinely achieved prior to gas introduction.

\subsection{Vacuum System}

Evacuation of both upstream and downstream sections is achieved using a Pfeiffer HiCube 80 NEO benchtop turbomolecular pump station equipped with an MVP 030-3 diaphragm roughing pump.
The pump station is connected to each section via KF40 stainless steel lines and all-metal valves, allowing independent evacuation and isolation.

The diaphragm pump provides initial roughing prior to turbomolecular operation, reducing system complexity and footprint.
Prior to hydrogen exposure, both sections are evacuated to high vacuum. 
Leak testing is performed after each sample installation to verify system integrity prior to operation.

\subsection{Pressure Measurement}

Pressure monitoring is performed using two complementary measurement systems in each section of the rig, as shown in Fig~\ref{fig:rig_schematic}.

Vacuum levels are monitored using InstruTech gauges: a CVM211 Stinger convection (Pirani) gauge and a WGM701 Wasp dual Pirani and cold cathode gauge.
The upstream vacuum gauge is labelled P1a, and the downstream vacuum gauge is labelled P2a.
These gauges provide wide-dynamic-range measurements during pump-down and leak checking.

Accurate hydrogen pressure measurements are obtained using capacitance manometers (Baratron 626D, unheated models).
The upstream driving pressure is measured using a \SI{1000}{Torr} full-scale gauge (P1b), while the downstream pressure rise is measured using a \SI{1}{Torr} full-scale gauge (P2b). 
Both gauges provide a resolution of 0.05\% of full scale, corresponding to detection limits of approximately \SI{0.5}{Torr} for P1b and \SI{5e-4}{Torr} for P2b. 
The use of capacitance manometers ensures gas-species-independent pressure measurement.

\subsection{Gas Injection System}

SHIELD operates using hydrogen at 99.999\% purity and argon at 99.99\% purity. 
Hydrogen is used for permeation measurements, while argon serves as a purge gas between experimental runs to minimise residual gas effects and cross-contamination.

Gas is supplied from cylinders through manual pressure regulators.
For each experiment, the upstream section is filled to the desired pressure, as measured by P1b, and subsequently isolated to establish a constant driving-pressure boundary condition.
The downstream section remains sealed during measurement.

\subsection{Sample Assembly}

Disc-shaped samples of \SI{20}{mm} diameter and \SI{1}{mm} thickness are mounted within 1/4~" Swagelok stainless steel VCR face-seal fittings.
This configuration provides high-integrity, repeatable sealing that is compatible with ultra-high-vacuum operation and thermal cycling.

The central region of the fitting accommodates the test specimen while maintaining mechanical stability and straightforward sample replacement between experimental runs.

\subsection{Heating System}

Temperature control is achieved with a Carbolite Gero TS split-tube furnace that encloses the central section of the VCR assembly.
This configuration ensures that the specimen is uniformly heated while adjacent piping remains near ambient temperature.

A type~K thermocouple is inserted through the helium leak-test port of the VCR fitting to maintain direct contact with the sample surface.
The thermocouple signal is continuously recorded, allowing direct measurement of sample temperature rather than relying solely on the furnace setpoint.
The operational temperature range of the platform is \SI{100}{\degreeCelsius} to \SI{600}{\degreeCelsius}.

\subsection{Data Acquisition and Control}

All pressure and temperature signals are recorded using a LabJack~U6 data acquisition system. 
Voltage outputs from P1a, P1b, P2a, P2b and the thermocouple are sampled and logged through an in-house Python-based framework.

The acquisition software, \texttt{SHIELD\_DAS}~\cite{dark_pttepxmitshield_das_2025}, is openly available on GitHub.
An associated automated data management system, \texttt{SHIELD-Data}~\cite{dark_shield-data_2025}, performs structured file organisation and automated backups to ensure data traceability.

The sampling frequency is configurable to \SI{5}{Hz}, with a typical acquisition interval of \SI{0.5}{s} used for permeation measurements.

\section{Results}
\label{sec:results}

The performance of the SHIELD platform was evaluated through hydrogen permeation measurements on stainless steel (316) and low-carbon steel (AISI 1018) specimens over a range of temperatures. 
The results demonstrate stable experimental conditions, reliable extraction of steady-state permeation flux, and agreement with established literature data within expected experimental variability.

\subsection{Permeation regime identification}

\begin{figure}[H]
    \centering
    \includegraphics[width=\linewidth, trim={0.6cm 0.6cm 0.6cm 0.6cm}, clip]{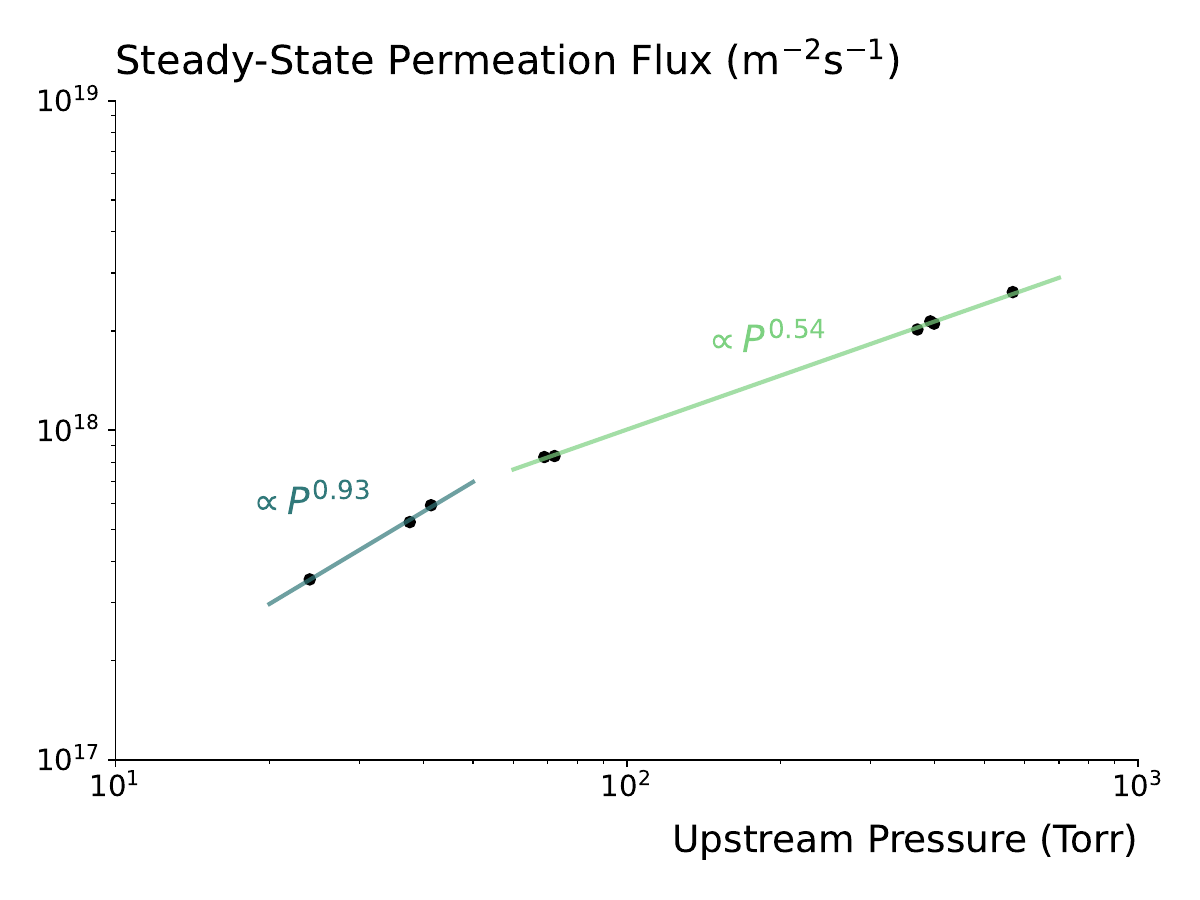}
    \caption{Permeation flux as a function of upstream hydrogen pressure for 316 stainless steel. At low pressures, the flux scales approximately linearly with pressure, indicating \textit{surface-limited} transport. At higher pressures, the dependence approaches a square-root scaling, consistent with \textit{diffusion-limited} permeation. The transition between regimes occurs at around \SI{80}{Torr} in the present experimental configuration.}
    \label{fig:measured_pressure_regimes}
\end{figure}

The dependence of permeation flux on upstream pressure was investigated to determine the dominant transport regime under the operating conditions of the SHIELD platform. 
The measured permeation flux as a function of upstream pressure for 316 stainless steel is shown in Fig.~\ref{fig:measured_pressure_regimes}.
As described in Section~\ref{sec:permeation_theory}, the governing transport regime can be identified from the scaling of permeation flux with pressure. 
In the \textit{surface-limited} regime, the flux is expected to scale linearly with pressure ($J \propto P$), whereas in the \textit{diffusion-limited} regime, the flux follows a square-root dependence ($J \propto \sqrt{P}$).
At lower pressures, the measured flux exhibits an approximately linear dependence on pressure, consistent with \textit{surface-limited} transport.
As the upstream pressure increases, a transition is observed towards a sub-linear dependence, approaching a square-root scaling with pressure, indicative of \textit{diffusion-limited} permeation.
This behaviour provides a direct experimental validation of the theoretical scaling relations outlined in Section~\ref{sec:permeation_theory}.
The transition between these regimes is observed at upstream pressures of approximately \SI{80}{Torr} in the present configuration. 
Above this pressure, the system operates predominantly in the \textit{diffusion-limited} regime, ensuring the validity of the steady-state analysis methods used to determine permeability.
These results confirm that the experimental conditions employed in this work are appropriate for extracting bulk transport properties and that surface kinetics effects do not dominate the measurements at higher pressures.

\subsection{Pressure rise measurement}

\begin{figure}[h]
    \centering
    \includegraphics[width=\linewidth, trim={0.7cm 0.7cm 0.7cm 0.5cm}, clip]{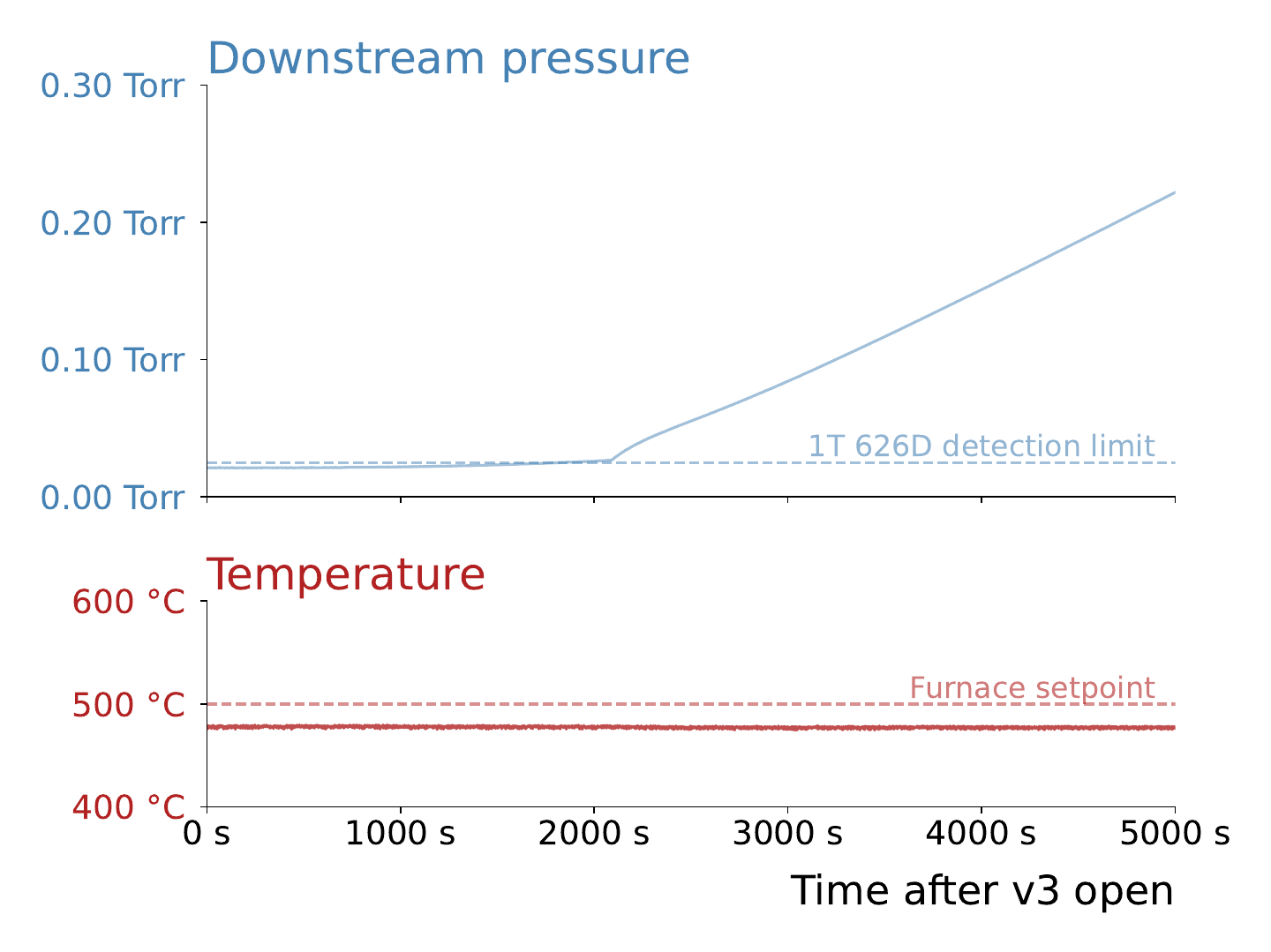}
    \caption{Downstream pressure evolution during a hydrogen permeation experiment on stainless steel, together with the corresponding sample temperature. Following an initial transient regime, a linear increase in downstream pressure is observed, indicating the establishment of steady-state permeation. The stable temperature profile confirms well-controlled thermal conditions during the measurement.}
    \label{fig:pressure_rise}
\end{figure}

A typical downstream pressure evolution from a permeation experiment is shown in Fig.~\ref{fig:pressure_rise}, together with the corresponding sample temperature.
Following the introduction of hydrogen into the upstream volume, the downstream pressure initially remains constant (at the lower detection limit of the pressure gauge) before increasing as hydrogen permeates through the specimen.

After an initial transient regime, the downstream pressure increases linearly, indicating the establishment of steady-state permeation. 
The stability of the sample temperature throughout the measurement confirms that thermal conditions are well controlled, ensuring consistent transport properties.
This behaviour is consistent with the expected response of a static gas-driven permeation system.

The clear linear regime demonstrates that the system provides sufficiently low noise and stable boundary conditions to accurately determine the steady-state permeation flux.

The early-time transient preceding steady-state permeation is only partially resolved in this measurement. 
This is primarily due to the full-scale range of the downstream capacitance manometer, which limits pressure resolution at very low pressures. 
Improved characterisation of the transient regime would require a lower full-scale pressure gauge to enhance sensitivity during the initial stages of permeation.
However, this limitation does not affect the determination of steady-state permeation flux, which is derived from the linear pressure rise at later times. 
The current configuration is therefore well suited to accurate permeability measurements, which constitute the primary objective of the SHIELD platform.

\subsection{Permeability measurements}

Hydrogen permeability was determined from the steady-state pressure rise in the downstream volume, as described in Section~\ref{sec:permeation_theory}. 
The steady-state regime was clearly identified in each experiment by the emergence of a linear increase in pressure, ensuring consistent extraction of the permeation flux across all measurements.
Once steady-state permeation is established, the downstream pressure increases linearly with time, and the permeation flux is obtained from the slope of this linear regime (see Equation \ref{eq:pressure-rise}).

\begin{figure}[h]
    \centering
    \includegraphics[width=\linewidth]{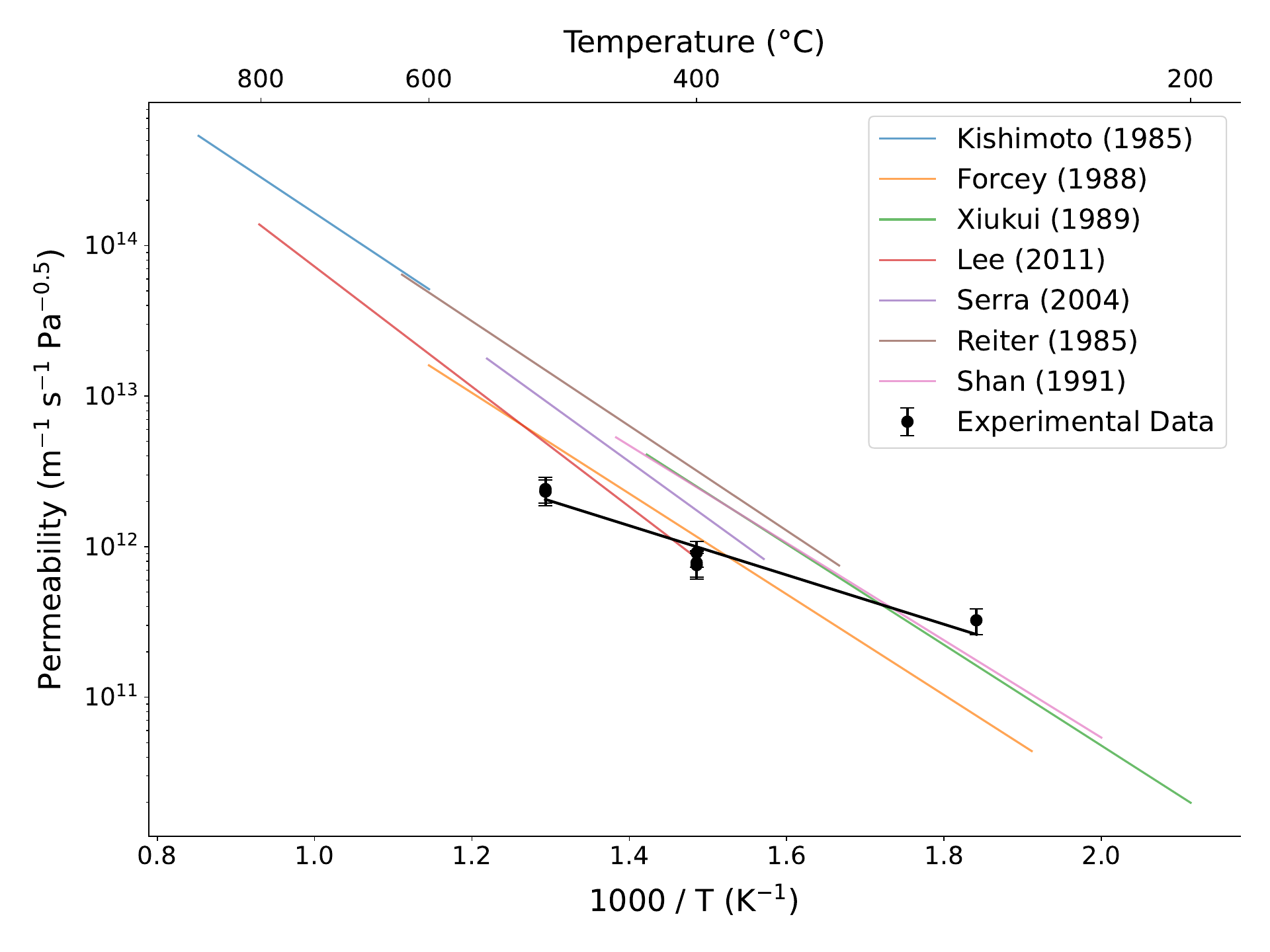}
    \caption{Hydrogen permeability measured using SHIELD as a function of inverse temperature for 316 stainless steel, compared with published literature~\cite{kishimoto_hydrogen_1985, forcey_hydrogen_1988, xiukui_hydrogen_1989, lee_hydrogen_2011, serra_hydrogen_2005-1, shan_behavior_1991, reiter_interaction_1985}. Literature values obtained from HTM~\cite{delaporte-mathurin_remdelaportemathurinh-transport-materials_2026}. Error bars represent propagated uncertainties from the pressure-based flux determination.}
    \label{fig:permeability_ss}
\end{figure}

The hydrogen permeability of 316 stainless steel measured using SHIELD is shown in Fig.~\ref{fig:permeability_ss} as a function of inverse temperature, alongside literature data~\cite{kishimoto_hydrogen_1985, forcey_hydrogen_1988, xiukui_hydrogen_1989, lee_hydrogen_2011, serra_hydrogen_2005-1, shan_behavior_1991, reiter_interaction_1985}.
Error bars on the measured permeability represent propagated uncertainties arising from determining the steady-state flux, based on pressure-measurement resolution and signal noise in the downstream pressure rise.
The spread observed in the literature data reflects known variability in material composition and surface condition, and the SHIELD measurements fall within this range.
The measured permeability exhibits a clear Arrhenius dependence, consistent with thermally activated transport.
The SHIELD measurements reproduce the general temperature dependence reported in the literature and show good agreement in magnitude within the variability of published datasets.
However, a slight deviation in slope relative to literature trends is observed, indicating a difference in the apparent activation behaviour. 
This may arise from differences in surface condition, oxide layer formation, or uncertainties in effective sample temperature, all of which can influence the measured permeation response.
The deviation suggests a modification of the apparent activation energy relative to literature values.
Further investigation is required to fully quantify the origin of this discrepancy.

\begin{figure}[H]
    \centering
    \includegraphics[width=\linewidth]{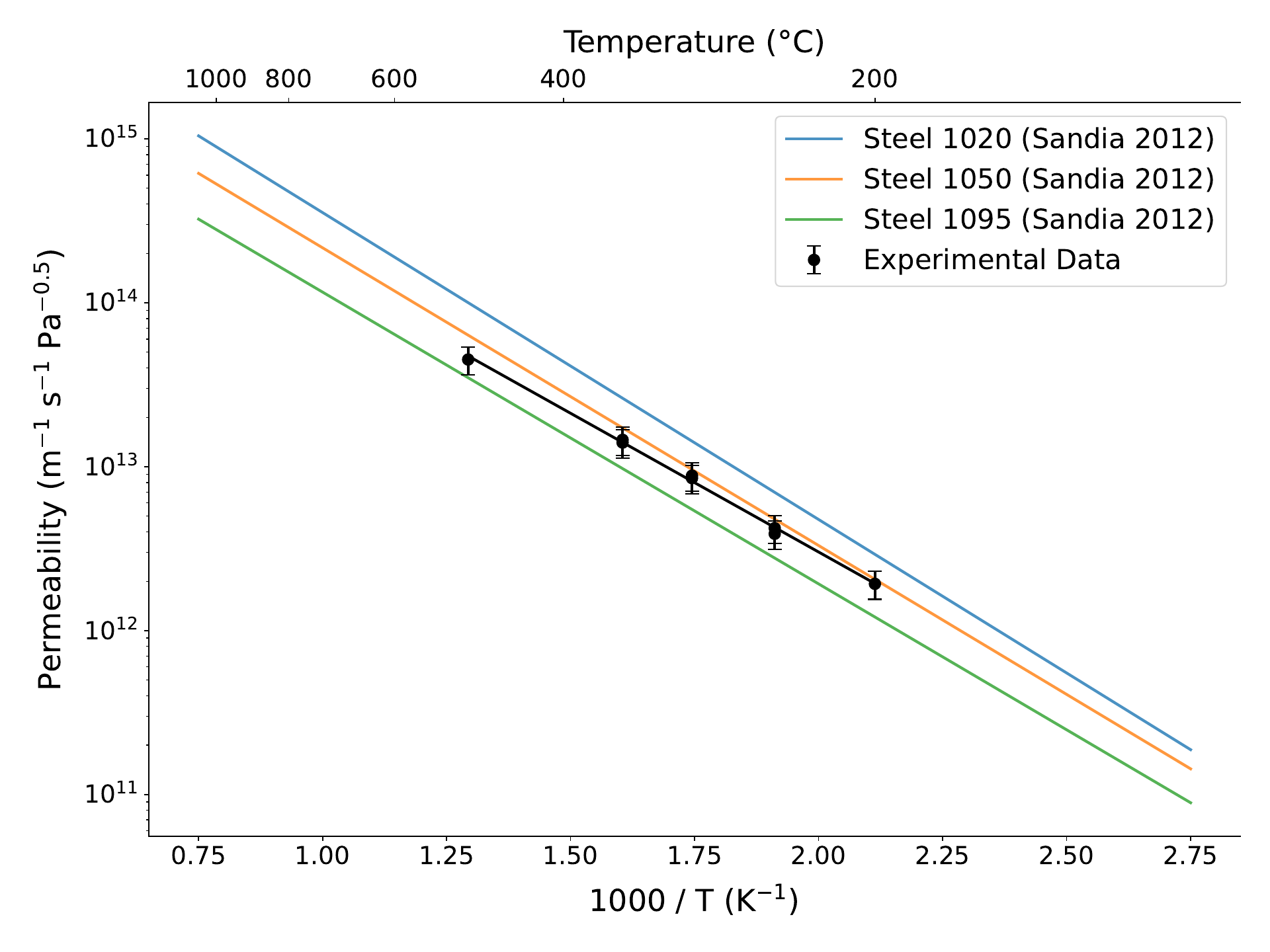}
    \caption{Hydrogen permeability measured using SHIELD as a function of inverse temperature for AISI 1018 low-carbon steel, compared with literature~\cite{sanmarchi2012hydrogen}. Literature values obtained from HTM~\cite{delaporte-mathurin_remdelaportemathurinh-transport-materials_2026}. Error bars represent propagated uncertainties from the pressure-based flux determination.}
    \label{fig:permeability_lc}
\end{figure}

Measurements performed on AISI 1018 low-carbon steel are presented in Fig.~\ref{fig:permeability_lc}, alongside literature data.
The measured permeability again exhibits Arrhenius behaviour and shows good agreement with published values over the investigated temperature range, both in magnitude and temperature dependence. 
In contrast to the stainless steel case, the temperature dependence closely follows the trends reported in the literature.
This consistency across different material systems demonstrates the robustness of the SHIELD platform and supports the reliability of the steady-state pressure-rise method.

Taken together, these results demonstrate that the SHIELD platform produces permeability measurements consistent with established datasets across a range of materials and conditions.

\subsection{Measurement uncertainty and limitations}

Minor deviations between the present measurements and literature values are observed, particularly at lower temperatures. 
These differences may arise from variations in material composition, surface condition, and oxide layer formation, all of which influence hydrogen transport and surface exchange kinetics.

Uncertainties in effective sample temperature and pressure measurement resolution may also contribute to systematic offsets in the extracted permeability.
The sample temperature is currently measured with a thermocouple inserted through the helium leak-test port of the VCR fitting, providing a close approximation of the specimen temperature. 
However, this configuration does not ensure direct contact with the sample surface, and small temperature gradients may exist between the thermocouple location and the specimen.
Such offsets in the effective sample temperature can influence the apparent activation behaviour and may contribute to the observed deviations in temperature dependence relative to literature data. 
Improved temperature measurement, for example by placing the thermocouple in direct contact with the sample surface, could reduce this uncertainty in future experiments.

Although diffusivity can, in principle, be determined using the time-lag method, its extraction was found to be sensitive to early-time pressure transients and baseline fluctuations in the present configuration. 
As a result, diffusivity values exhibited significant variability and are not reported here. 

In contrast, steady-state measurements provide a more robust and reproducible determination of permeability, which is the primary quantity of interest for permeation barrier evaluation.
Despite these limitations, the permeability measurements obtained with SHIELD are consistent and reproducible, supporting the platform's validity for hydrogen permeation studies.

\section{Outlook and future work}
\label{sec:outlook}

The SHIELD gas-driven permeation rig has been developed as a dedicated experimental platform for investigating hydrogen transport and permeation barrier performance under controlled thermal and pressure conditions. 
Having established stable operation and reproducible measurement workflows, the next phase of work will focus on systematic evaluation of coated systems.

Initial studies will examine tungsten coatings deposited by physical vapour deposition and silicon carbide coatings deposited by chemical vapour deposition.
These materials represent candidate barrier systems for high-temperature structural applications. 
For highly effective coatings, the resulting steady-state permeation flux may approach the current downstream detection threshold.
In such cases, extended-duration experiments or improved pressure resolution will be required to quantify permeability accurately.
Establishing the measurable performance envelope of SHIELD for low-permeability systems will therefore form an integral part of upcoming investigations.

A major planned development is the introduction of multi-isotope permeation capability using hydrogen and deuterium.
Integration of a quadrupole mass spectrometer into the downstream section will enable isotope-resolved detection of H$_2$, D$_2$, and HD.
This will allow the direct separation of isotopic contributions to the permeation signal and provide access to surface recombination and isotope-exchange behaviour.
Such measurements will extend the platform beyond conventional steady-state permeability determination, enabling experimental investigation of recombination kinetics and non-ideal surface effects under mixed-isotope conditions.

The enhanced sensitivity of mass spectrometry may significantly expand the detectable permeability range for the rig, particularly for advanced barrier systems.
However, implementation will require careful modification of the downstream vacuum configuration to meet the mass spectrometer's operational requirements while preserving the well-defined boundary conditions characteristic of static GDP experiments.

In parallel, SHIELD will support the evaluation of permeation barrier coatings intended for use in molten-salt environments.
While detailed salt-exposure studies will be reported separately, the present platform provides controlled hydrogen-transport measurements necessary to characterise coating performance before environmental testing and to assess potential degradation following thermal cycling.

\section{Conclusions}
The SHIELD gas-driven permeation platform has been developed as a dedicated experimental system for measuring hydrogen transport properties under controlled thermal and pressure conditions. 
The rig provides stable boundary conditions, reliable pressure measurement over a wide dynamic range, and reproducible permeation measurements.

Analysis of the pressure dependence of permeation flux confirms that the system operates in the diffusion-limited regime above approximately \SI{80}{Torr}, with \SI{1}{mm} 316 stainless steel samples, thereby ensuring the validity of the steady-state methods used to determine permeability.

Measurements performed on 316 stainless steel and AISI 1018 low-carbon steel show good agreement with established literature data across the investigated temperature range. 
The measured permeability exhibits Arrhenius behaviour and falls within the spread of reported values, demonstrating that the platform can deliver quantitatively reliable permeation data across different material systems.

The present work also identifies practical limitations associated with the current configuration. 
Although diffusivity can in principle be obtained from transient permeation behaviour, the values extracted here were not sufficiently reproducible to be reported with confidence. 
Uncertainties in effective sample temperature are a likely contributing factor. 
By contrast, steady-state permeability measurements remain robust and reproducible.

Overall, SHIELD provides a well-characterised and flexible platform for hydrogen permeation experiments. 
The combination of controlled operating conditions, straightforward sample integration, and an openly documented data acquisition framework makes the system well-suited for routine permeability measurements and for the evaluation of permeation barrier coatings.


 \section*{References}
\bibliography{references}

\end{document}